\documentstyle[amsfonts,eqsecnum,aps]{revtex}

\begin{document}
\draft
\title{Density and Pair Correlation Function of Confined Identical Particles: the
Bose-Einstein Case}
\author{F.~Brosens and J.T. Devreese\cite{Author2}}
\address{Departement Natuurkunde, Universiteit Antwerpen (UIA),\\
Universiteitsplein 1, B-2610 Antwerpen}
\author{L. F. Lemmens}
\address{Departement Natuurkunde, Universiteit Antwerpen (RUCA),\\
Groenenborgerlaan 171, B-2020 Antwerpen}
\date{January 6,1997; revised version March 7, 1997.}
\maketitle

\begin{abstract}
Two basic correlation functions are calculated for a model of $N$
harmonically interacting identical particles in a parabolic potential well.
The density and the pair correlation function of the model are investigated
for the boson case. The dependence of these static response properties on
the complete range of the temperature and of the number of particles is
obtained. The calculation technique is based on the path integral approach
of symmetrized density matrices for identical particles in a parabolic
confining well.
\end{abstract}


\section{Introduction}

Generalizing the Feynman approach of identical particles in a box \cite
{Feynman} to the case of identical particles in a parabolic confining
potential, the present authors derived analytic expressions for the
propagator and for the partition function of a system of $N$ harmonically
interacting {\sl identical} particles (bosons or fermions) in a parabolic
well \cite{BDLprevious}, hereafter referred to as I. This model, giving rise
to repetitive Gaussian integrals, also allows to obtain the generating
function for the partition function. For an ideal gas of non-interacting
particles in a parabolic well, this generating function coincides with the
grand-canonical partition function. For interacting particles this
generating function circumvents the constraints on the summation over the
cycles of the permutation group at the expense of doing an extra path
integral.

In the present paper the one-point and two-point correlation functions of
the model are calculated using their generating function as we did for the
thermodynamic properties of the model. Also here we have to introduce extra
path integrals of Gaussian nature to facilitate the cyclic summations.

The one-point and two-point correlation functions of the model are obtained
for the boson case as well as for the fermion case. But in view of the
recent interest in Bose-Einstein condensation in a trap \cite{BEC1,BEC2,BEC3}%
, the explicit evaluation and the discussion of the results are restricted
to the boson case in the present paper. The fermion case will be studied in
a forthcoming paper.

In the case of distinguishable particles, the correlation functions play a
key role in the variational approximation for path integrals \cite
{FeynHibbs,FeynKlei}. This variational method can be reformulated for
indistinguishable particles, and the knowledge of the one- and two-point
correlation functions for harmonic trial actions is as crucial as it is for
distinguishable particles. For any algorithmic approach to many-body
diffusion \cite{LBD,BDL,LBDPR} for interacting particles, the knowledge of
the correlation functions of the model is very useful to test the actual
implementations. Furthermore, the model provides an example of an exactly
tractable system with interactions, which clearly exhibits the effects of
Bose-Einstein condensation in the specific heat \cite{BDLheat} and in the
moment of inertia \cite{BDLinertia}.

The one-body potential energy $V_1$ and the two-body potential energy $V_2$
of the model system are given by
\begin{equation}
V=V_1+V_2;\quad V_1=\frac{m\Omega ^2}2\sum_{j=1}^N\left. {\bf r}_j\right.
^2;\quad V_2=-\frac{m\omega ^2}4\sum_{j,l=1}^N\left( {\bf r}_j-{\bf r}%
_l\right) ^2.
\end{equation}
The two-body interaction is assumed to be repulsive; replacing $-\omega ^2$
by $\omega ^2$ in $V_2$ gives the attractive case. As a result of the
diagonalization one obtains {\sl in each dimension} one degree of freedom
(the center of mass) with frequency $\Omega ,$ and $N-1$ degrees of freedom
with frequency $w$ given by
\begin{equation}
w=\sqrt{\Omega ^2-N\omega ^2}.
\end{equation}
For this many-body system distinguishability of the particles therefore
implies that one is dealing with a system that reduces to $3N$ degrees of
freedom, each degree of freedom representing one linear harmonic oscillator.
It is clear that for such a system the propagator, the thermodynamic
functions and the correlation functions are well known \cite{FordKacMazur}.

For identical particles (bosons or fermions) the propagator can be obtained
from the decomposition of the underlying processes in terms of 4 orthogonal
processes with well-defined boundary conditions \cite{LBD,BDL,LBDPR}. A
typical sample path for fermions is provided by the subprocess with
absorption at the boundary for the $x$-direction (leading to a fermion
diffusion process), while for the $y$- and the $z$-direction a boson
diffusion process with reflection at the boundary has to be used. This
procedure is used to generate the trajectory of the walkers (including the $%
y $- and $z$--components). The path of a walker in this particular
subprocess terminates if the motion in euclidean time gets absorbed along
the $x$-direction. Indistinguishability has therefore the important effect
of making the coupled oscillator problem in $3$ dimensions also a genuine 3D
problem.

In I we showed how extra degrees of freedom for the center of mass can be
introduced (in Fourier space) leading to a propagator which factorizes. The
extra degrees of freedom can be integrated out afterwards. As a consequence,
the internal degrees of freedom of the interacting oscillator system can be
considered as the degrees of freedom of another non-interacting oscillator
system. This mapping allows to use the grand-canonical partition function of
the non-interacting system as the generating function for the system in
interaction, provided the fugacity and hence the thermodynamical potential
are identified as usual.

The calculation of the density and the pair correlation function heavily
relies on the calculations presented in I, which makes it difficult to make
this paper self-contained without repeating some of the material presented
in I. We tried to partly overcome this inconvenience by using the same
notation as in I, and by a limited number of explicit references to the
detailed manipulations in I if similar situations are encountered.

The paper is organized as follows. In the next section the one- and
two-point correlation functions are calculated for identical particles
(bosons or fermions). In the third section the density and the pair
correlation function are analyzed for the boson case. In the fourth section
we discuss the results and draw some conclusions.

\section{Static response properties of a many-body system}

For the calculation of the static response properties of a many-body system,
the correlation functions $\sum_l\left\langle e^{i{\bf q}\cdot {\bf r}%
_l}\right\rangle _I$ and $\sum_{l,l^{\prime }}\left\langle e^{i{\bf q}\cdot
\left( {\bf r}_l-{\bf r}_{l^{\prime }}\right) }\right\rangle _I$ are the key
ingredients. The subscript $I$ emphasizes that identical particles are
considered (which can be specified to be bosons with subscript $B$ or
fermions with subscript $F$) in 3D. In the path integral approach the
expectation values of an expression $A\left( {\bf \bar{r}}^{\prime },\tau
\right) $ are given by
\begin{equation}
\left\langle A\left( {\bf \bar{r}}^{\prime },\tau \right) \right\rangle _I=%
\frac{\int d{\bf \bar{r}}\int d{\bf \bar{r}}^{\prime }K_I\left( {\bf \bar{r}}%
,\beta |{\bf \bar{r}}^{\prime },\tau \right) A\left( {\bf \bar{r}}^{\prime
},\tau \right) K_I\left( {\bf \bar{r}}^{\prime },\tau |{\bf \bar{r}}%
,0\right) }{\int d{\bf \bar{r}}K_I\left( {\bf \bar{r}},\beta |{\bf \bar{r}}%
,0\right) },
\end{equation}
where $K_I$ is the statistical propagator of the identical particles and $%
{\bf \bar{r}}$ is the $3N$-dimensional vector containing the coordinates $%
{\bf r}_1,\cdots ,{\bf r}_N$ of the $N$ particles. In this notation the
probability density, the pair correlation function and their Fourier
transforms are given by
\begin{equation}
n\left( {\bf r}\right) =\frac 1N\left\langle \sum_{l=1}^N\delta \left( {\bf %
r-r}_l\right) \right\rangle _I=\int \frac{d{\bf q}}{\left( 2\pi \right) ^3}%
n_{{\bf q}}e^{-i{\bf q\cdot r}}\longleftrightarrow n_{{\bf q}}=\frac 1N%
\sum_{l=1}^N\left\langle e^{i{\bf q}\cdot {\bf r}_l}\right\rangle _I,
\end{equation}
\begin{equation}
g\left( {\bf r}\right) =\frac 1N\left\langle \sum_{l,l^{\prime }\neq
l}^N\delta \left( {\bf r-r}_l+{\bf r}_{l^{\prime }}\right) \right\rangle
_I=\int \frac{d{\bf q}}{\left( 2\pi \right) ^3}g_{{\bf q}}e^{-i{\bf q\cdot r}%
}\longleftrightarrow g_{{\bf q}}=\frac 1N\sum_{l,l^{\prime }\neq
l}^N\left\langle e^{i{\bf q}\cdot \left( {\bf r}_l-{\bf r}_{l^{\prime
}}\right) }\right\rangle _I.
\end{equation}

Collecting the appropriate expressions for the propagators $K_I\left( {\bf
\bar{r}},\beta |{\bf \bar{r}}^{\prime },\tau \right) $ and $K_I\left( {\bf
\bar{r}}^{\prime },\tau |{\bf \bar{r}},0\right) $ from I, one sees that the
Fourier transforms $n_{{\bf q}}$ and $g_{{\bf q}}$ are given by
\begin{equation}
n_{{\bf q}}=\frac 1{NZ_I}\int \int \frac{d{\bf R}d{\bf k}}{\left( 2\pi
\right) ^3}e^{i{\bf k}\cdot {\bf R}}\int d{\bf \bar{r}}e^{-i{\bf \bar{k}}%
\cdot {\bf \bar{r}}}\sum_le^{i{\bf q}\cdot {\bf r}_l}\frac 1{N!}\sum_p\xi
^p\prod_{j=1}^NK\left( \left( P{\bf r}\right) _j,\beta |{\bf r}_j,0\right)
_w,
\end{equation}
\begin{equation}
g_{{\bf q}}=\frac 1{NZ_I}\int \int \frac{d{\bf R}d{\bf k}}{\left( 2\pi
\right) ^3}e^{i{\bf k}\cdot {\bf R}}\int d{\bf \bar{r}}e^{-i{\bf \bar{k}}%
\cdot {\bf \bar{r}}}\frac 1{N!}\sum_p\xi ^p\sum_{l,l^{\prime }\neq l}^Ne^{i%
{\bf q}\cdot \left( {\bf r}_l-{\bf r}_{l^{\prime }}\right)
}\prod_{j=1}^NK\left( \left( P{\bf r}\right) _j,\beta |{\bf r}_j,0\right) _w,
\end{equation}
where $Z_I$ is the partition function, $K\left( {\bf r}_j^{\prime },\beta |%
{\bf r}_j,0\right) _w$ is the propagator of a 3D harmonic oscillator with
frequency $w,$ $P$ denotes a permutation matrix and $\xi =-1$ assures the
required anti-symmetry for fermions, whereas $\xi =+1$ describes bosons.

We first show how a tractable expression can be obtained for $n_{{\bf q}},$
and subsequently use an analogous procedure to calculate $g_{{\bf q}}$.

\subsection{The single-particle expectation values}

Using the cyclic decomposition, and denoting by $M_\ell $ the number of
cycles of length$~\ell ,$ the expression for $n_{{\bf q}}$ can be written in
terms of the cycles in the same way as we did for the partition function in
I:
\begin{eqnarray}
n_{{\bf q}} &=&\frac 1{NZ_I}\int \int \frac{d{\bf R}d{\bf k}}{\left( 2\pi
\right) ^3}e^{i{\bf k}\cdot {\bf R}}  \nonumber \\
&&\times \sum_{M_1\cdots M_N}\sum_\ell \ell M_\ell {\cal K}_\ell \left( {\bf %
k},{\bf q}\right) \frac{\xi ^{\left( \ell -1\right) M_\ell }}{M_\ell !\ell
^{M_\ell }}\left( {\cal K}_\ell \left( {\bf k}\right) \right) ^{M_\ell
-1}\prod_{\ell ^{\prime }\neq \ell }\frac{\xi ^{\left( \ell ^{\prime
}-1\right) M_{\ell ^{\prime }}}}{M_{\ell ^{\prime }}!\left( \ell ^{\prime
}\right) ^{M_{\ell ^{\prime }}}}\left( {\cal K}_{\ell ^{\prime }}\left( {\bf %
k}\right) \right) ^{M_{\ell ^{\prime }}},
\end{eqnarray}
where
\begin{equation}
{\cal K}_\ell \left( {\bf k},{\bf q}\right) =\int d{\bf r}_{\ell +1}\int d%
{\bf r}_\ell \cdots \int d{\bf r}_1\delta \left( {\bf r}_{\ell +1}-{\bf r}%
_1\right) e^{i{\bf q}\cdot {\bf r}_1}\prod_{j=1}^\ell K\left( {\bf r}%
_{j+1},\beta |{\bf r}_j,0\right) _we^{-i\frac 1N{\bf k\cdot r}_j},
\end{equation}
and ${\cal K}_\ell \left( {\bf k}\right) ={\cal K}_\ell \left( {\bf k},{\bf q%
}=0\right) $ is precisely the same function as found in Eq. (2.20) of I for
the determination of the partition function. Both the integrations over $%
{\bf k}$ and ${\bf R}$ only involve Gaussian integrands, and one eventually
finds after some algebra
\begin{equation}
n_{{\bf q}}=\exp \left[ -\frac{\hbar q^2}{4mN}\left( \frac{\coth \frac 12%
\beta \hbar \Omega }\Omega -\frac{\coth \frac 12\beta \hbar w}w\right)
\right] \tilde{n}_{{\bf q}}.
\end{equation}

The factor in front of $\tilde{n}_{{\bf q}}$ accounts for the center of
mass, and obviously reduces to unity in the non-interacting case where $%
w=\Omega .$ The factor $\tilde{n}_{{\bf q}}$ itself denotes the expectation
value of $\sum_le^{i{\bf q}\cdot {\bf r}_l}$ in the subspace of the {\sl %
relative coordinate system only} with partition function ${\Bbb Z}_I\left(
N\right) $:
\begin{equation}
\tilde{n}_{{\bf q}}=\frac 1{N{\Bbb Z}_I\left( N\right) }\sum_{M_1\cdots
M_N}\left[ \sum_\ell M_\ell \ell \exp \left( -\frac{\hbar q^2\coth \frac 12%
\ell \beta \hbar w}{4mw}\right) \right] \prod_\ell \frac{\xi ^{\left( \ell
-1\right) M_\ell }}{M_\ell !\ell ^{M_\ell }}\left( \frac 1{2\sinh \frac 12%
\ell \beta \hbar w}\right) ^{3M_\ell }.
\end{equation}

In ${\cal K}_\ell \left( {\bf k},{\bf q}\right) $ one recognizes the
partition function (over a time interval $\ell \beta $) of a driven 3D
harmonic oscillator with frequency $w$
\begin{equation}
{\cal K}_\ell \left( {\bf k},{\bf q}\right) =\int d{\bf r}K\left( {\bf r}%
,\ell \beta |{\bf r},0\right) _we^{-\int_0^{\ell \beta }d\tau {\bf f}_{{\bf q%
}}\left( \tau \right) \cdot {\bf r}\left( \tau \right) };\quad {\bf f}_{{\bf %
q}}\left( \tau \right) =i\frac 1N{\bf k}\sum_{j=1}^{\ell -1}\delta \left(
\tau -j\beta \right) +i\left( \frac{{\bf k}}N-{\bf q}\right) \delta \left(
\tau \right) ,
\end{equation}
which is known \cite{Feynman,FeynHibbs} in closed form:
\begin{equation}
{\cal K}_\ell \left( {\bf k},{\bf q}\right) =\frac 1{\left( 2\sinh \frac{%
\ell \beta }2\hbar w\right) ^3}e^{\Phi _{{\bf q}}};\quad \Phi _{{\bf q}}=%
\frac \hbar 2\int_0^{\ell \beta }d\tau \int_0^{\ell \beta }d\sigma \frac{%
{\bf f}_{{\bf q}}\left( \tau \right) \cdot {\bf f}_{{\bf q}}\left( \sigma
\right) }{2mw}\frac{\cosh \left( \frac{\ell \beta }2-\left| \tau -\sigma
\right| \right) \hbar w}{\sinh \frac 12\ell \beta \hbar w}.
\end{equation}
The calculation of $\Phi _{{\bf q}},$ given $f_{{\bf q}}\left( \tau \right) $
as a sum of delta-functions, is straightforward. The result is
\begin{equation}
\Phi _{{\bf q}}=-\frac \hbar {4mw}\left( \frac{\ell k^2}{N^2}\coth \frac 12%
\beta \hbar w-2\frac{{\bf k}\cdot {\bf q}}N\coth \frac 12\beta \hbar
w+q^2\coth \frac 12\ell \beta \hbar w\right) ,
\end{equation}
and consequently:
\begin{eqnarray}
{\cal K}_\ell \left( {\bf k},{\bf q}\right) &=&{\cal K}_\ell \left( {\bf k}%
\right) \exp \left( \frac{\hbar {\bf k}\cdot {\bf q}}{2Nmw}\coth \frac 12%
\beta \hbar w-\frac{\hbar q^2}{4mw}\coth \frac 12\ell \beta \hbar w\right) ,
\\
{\cal K}_\ell \left( {\bf k}\right) &=&\frac 1{\left( 2\sinh \frac{\ell
\beta }2\hbar w\right) ^3}\exp \left( -\frac \hbar {4mw}\frac{\ell k^2}{N^2}%
\coth \frac 12\beta \hbar w\right) .
\end{eqnarray}

Introducing the generating function ${\cal G}_1\left( u,{\bf q}\right)
=\sum_{N=0}^\infty \left[ {\Bbb Z}_I\left( N\right) N\tilde{n}_{{\bf q}%
}\right] u^N$:
\begin{equation}
{\cal G}_1\left( u,{\bf q}\right) =\sum_{N=0}^\infty \sum_{M_1\cdots
M_N}\left[ \sum_\ell M_\ell \ell \exp \left( -\frac{\hbar q^2\coth \frac 12%
\ell \beta \hbar w}{4mw}\right) \right] \prod_\ell \frac 1{M_\ell !}\left[
\frac{\xi ^{\left( \ell -1\right) }u^\ell }{\ell \left( 2\sinh \frac 12\ell
\beta \hbar w\right) ^3}\right] ^{M_\ell },
\end{equation}
the summations can be done:
\begin{equation}
{\cal G}_1\left( u,{\bf q}\right) =\Xi _I\left( u\right) \sum_{\ell
=1}^\infty \frac{\xi ^{\ell -1}\exp \left( -\frac{\hbar q^2}{4mw}\coth \frac %
12\ell \beta \hbar w\right) }{\left( 2\sinh \frac 12\ell \beta \hbar
w\right) ^3}u^\ell ,
\end{equation}
where $\Xi _I\left( u\right) =\sum_{N=0}^\infty {\Bbb Z}_I\left( N\right)
u^N $ is the generating function of the partition function ${\Bbb Z}_I\left(
N\right) $ of $N$ identical oscillators in the relative coordinate system,
studied in I. Consequently
\begin{equation}
\tilde{n}_{{\bf q}}=\frac 1N\sum_{\ell =1}^N\frac{\xi ^{\ell -1}\exp \left( -%
\frac{\hbar q^2}{4mw}\coth \frac 12\ell \beta \hbar w\right) }{\left( 2\sinh
\frac 12\ell \beta \hbar w\right) ^3}\frac{{\Bbb Z}_I\left( N-\ell \right) }{%
{\Bbb Z}_I\left( N\right) }.
\end{equation}
Considering the limit ${\bf q\rightarrow }0{\bf ,}$ it should be noted that
the sum rule $\tilde{n}_{{\bf q}=0}=1$ is indeed satisfied.

\subsection{The two-particle expectation values}

Similarly as for the treatment of the single-particle correlation function,
the Fourier transform which allows to treat the center-of-mass coordinate as
an independent degree of freedom is introduced. The cyclic decomposition of
the permutations implies that a factor $e^{i{\bf q}\cdot {\bf r}_l}$ occurs
once in each position of each cycle. Furthermore a different factor $e^{i%
{\bf q}\cdot {\bf r}_{l^{\prime }}}$ occurs in each position of each cycle
which differs from ${\bf r}_l$ (i.e. the case $l=l^{\prime }$ has to be
excluded if ${\bf r}_l$ and ${\bf r}_{l^{\prime }}$ are within the {\sl same}
cycle). Taking these bookkeeping considerations into account, the cyclic
decomposition of the summation over the permutations leads to
\begin{eqnarray}
g_{{\bf q}} &=&{\frac 1{Z_IN}}\int \int \frac{d{\bf R}d{\bf k}}{\left( 2\pi
\right) ^3}e^{i{\bf k}\cdot {\bf R}}\sum_{M_1\cdots M_N}\left( \prod_\ell
\frac{\xi ^{\left( \ell -1\right) M_\ell }}{M_\ell !\ell ^{M_\ell }}\right)
\times  \nonumber \\
&&\times \sum_\ell \ell M_\ell \left(
\begin{array}{l}
\sum_{j=1}^{\ell -1}{\cal K}_\ell \left( {\bf k},{\bf q;}j+1\right) \left(
{\cal K}_\ell \left( {\bf k}\right) \right) ^{M_\ell -1}\prod_{\ell ^{\prime
}\neq \ell }\left( {\cal K}_{\ell ^{\prime }}\left( {\bf k}\right) \right)
^{M_{\ell ^{\prime }}} \\
+\ell \left( M_\ell -1\right) {\cal K}_\ell \left( {\bf k},{\bf q}\right)
{\cal K}_\ell \left( {\bf k},-{\bf q}\right) \left( {\cal K}_\ell \left(
{\bf k}\right) \right) ^{M_\ell -2}\prod_{\ell ^{\prime }\neq \ell }\left(
{\cal K}_{\ell ^{\prime }}\left( {\bf k}\right) \right) ^{M_{\ell ^{\prime
}}} \\
+\sum_{\ell ^{\prime }\neq \ell }\ell ^{\prime }M_{\ell ^{\prime }}{\cal K}%
_\ell \left( {\bf k},{\bf q}\right) {\cal K}_{\ell ^{\prime }}\left( {\bf k}%
,-{\bf q}\right) \left( {\cal K}_\ell \left( \vec{k}\right) \right) ^{M_\ell
-1}\left( {\cal K}_\ell \left( {\bf k}\right) \right) ^{M_{\ell ^{\prime
}}-1}\prod_{\ell ^{\prime \prime }\neq \ell ,\ell ^{\prime }}\left( {\cal K}%
_{\ell ^{\prime \prime }}\left( {\bf k}\right) \right) ^{M_{\ell ^{\prime
\prime }}}
\end{array}
\right) .
\end{eqnarray}
where ${\cal K}_\ell \left( {\bf k},{\bf q}\right) $ and ${\cal K}_\ell
\left( {\bf k}\right) $ are defined as in the previous subsection, and a
function ${\cal K}_\ell \left( {\bf k},{\bf q;}j\right) $ is introduced
which is given by
\begin{equation}
{\cal K}_\ell \left( {\bf k},{\bf q;}j\right) =\int d{\bf r}_{\ell +1}\int d%
{\bf r}_\ell \cdots \int d{\bf r}_1\delta \left( {\bf r}_{\ell +1}-{\bf r}%
_1\right) e^{i{\bf q}\cdot {\bf r}_1}e^{-i{\bf q}\cdot {\bf r}%
_j}\prod_{j^{\prime }=1}^\ell K\left( {\bf r}_{j^{\prime }+1},\beta |{\bf r}%
_{j^{\prime }},0\right) _we^{-i\frac 1N{\bf k\cdot r}_{j^{\prime }}}.
\end{equation}
In ${\cal K}_\ell \left( {\bf k},{\bf q;}j\right) $ one recognizes the
partition function (over a time interval $\ell \beta $) of a driven 3D
harmonic oscillator with frequency $w$%
\begin{equation}
{\cal K}_\ell \left( {\bf k},{\bf q;}j+1\right) =\int d{\bf r}K\left( {\bf r}%
,\ell \beta |{\bf r},0\right) _we^{-\int_0^{\ell \beta }d\tau {\bf h}_{{\bf q%
}}\left( \tau ,j\right) \cdot {\bf r}\left( \tau \right) };\quad {\bf h}_{%
{\bf q}}\left( \tau ,j\right) =i\frac 1N{\bf k}\sum_{j^{\prime }=0}^{\ell
-1}\delta \left( \tau -j^{\prime }\beta \right) -i{\bf q}\delta \left( \tau
\right) +i{\bf q}\delta \left( \tau -j\beta \right) .
\end{equation}
Similarly as for the single-particle correlation functions this expression
is known \cite{Feynman,FeynHibbs} in closed form:
\begin{equation}
{\cal K}_\ell \left( {\bf k},{\bf q;}j+1\right) =\left( \frac 1{2\sinh \frac{%
\ell \beta }2\hbar w}\right) ^3e^{\Psi _{{\bf q}}\left( j\right) };\quad
\Psi _{{\bf q}}\left( j\right) =\frac \hbar 2\int_0^{\ell \beta }d\tau
\int_0^{\ell \beta }d\sigma \frac{{\bf h}_{{\bf q}}\left( \tau ,j\right)
\cdot {\bf h}_{{\bf q}}\left( \tau ,j\right) }{2mw}\frac{\cosh \left( \frac{%
\ell \beta }2-\left| \tau -\sigma \right| \right) \hbar w}{\sinh \frac 12%
\ell \beta \hbar w}.
\end{equation}

The explicit evaluation of the influence function $\Psi _{{\bf q}}\left(
j\right) $ is somewhat involved but straightforward:
\begin{equation}
\Psi _{{\bf q}}\left( j\right) =-\frac \ell {N^2}\frac{\hbar {\bf k}^2}{4mw}%
\frac{e^{\beta \hbar w}+1}{e^{\beta \hbar w}-1}-\frac{\hbar {\bf q}^2}{2mw}%
\frac{\cosh \frac 12\ell \beta \hbar w-\cosh \left( \frac 12\ell -j\right)
\beta \hbar w}{\sinh \frac 12\ell \beta \hbar w},
\end{equation}
and consequently
\begin{equation}
{\cal K}_\ell \left( {\bf k},{\bf q;}j+1\right) ={\cal K}_\ell \left( {\bf k}%
\right) \exp \left( -\frac{\hbar q^2}{2mw}\frac{\cosh \frac 12\ell \beta
\hbar w-\cosh \left( \frac 12\ell -j\right) \beta \hbar w}{\sinh \frac 12%
\ell \beta \hbar w}\right) .
\end{equation}

Using the results obtained in the previous subsection for ${\cal K}_\ell
\left( {\bf k},{\bf q}\right) $ and ${\cal K}_\ell \left( {\bf k}\right) ,$
the Fourier transform of the pair correlation function becomes:
\begin{eqnarray}
g_{{\bf q}} &=&{\frac 1{Z_IN}}\int \int \frac{d{\bf R}d{\bf k}}{\left( 2\pi
\right) ^3}e^{i{\bf k}\cdot {\bf R}}\times \sum_{M_1\cdots M_N}\left(
\prod_\ell \frac{\xi ^{\left( \ell -1\right) M_\ell }}{M_\ell !\ell ^{M_\ell
}}\right) \prod_\ell \left( {\cal K}_\ell \left( {\bf k}\right) \right)
^{M_\ell }  \nonumber \\
&&\times \sum_\ell \ell M_\ell \left(
\begin{array}{l}
\sum_{j=1}^{\ell -1}\exp \left( -\frac{\hbar q^2}{2mw}\frac{\cosh \frac 12%
\ell \beta \hbar w-\cosh \left( \frac 12\ell -j\right) \beta \hbar w}{\sinh
\frac 12\ell \beta \hbar w}\right) \\
+\ell \left( M_\ell -1\right) \exp \left( -\frac 12\frac{\hbar q^2}{mw}\coth
\frac 12\ell \beta \hbar w\right) \\
+\sum_{\ell ^{\prime }\neq \ell }\ell ^{\prime }M_{\ell ^{\prime }}\exp
\left( -\frac 14\frac{\hbar q^2}{mw}\left( \coth \frac 12\ell \beta \hbar
w+\coth \frac 12\ell ^{\prime }\beta \hbar w\right) \right)
\end{array}
\right) .
\end{eqnarray}

The condition $\sum_\ell \ell M_\ell =N$ on the cyclic decompositions
simplifies $\prod_\ell \left( {\cal K}_\ell \left( {\bf k}\right) \right)
^{M_\ell }$, and the integrations over ${\bf k}$ and ${\bf R}$ are then
straightforward, resulting in $\left( \frac{\sinh \beta \hbar w/2}{\sinh
\beta \hbar \Omega /2}\right) ^3.$ The remaining summation over the cycles
can again be done if one introduces the appropriate generating function
\begin{equation}
{\cal G}_2\left( u,{\bf q}\right) =\sum_{N=0}^\infty \left[ {\Bbb Z}_I\left(
N\right) Ng_{{\bf q}}\right] u^N.
\end{equation}
which lifts the restriction on the number of cycles of given length. The
summation is straightforward. With $b=e^{-\beta \hbar w}$ one obtains:
\begin{equation}
{\cal G}_2\left( u,{\bf q}\right) =\Xi _I\left( u\right) \left\{ \sum_{\ell
=1}^\infty \frac{\xi ^{\ell -1}u^\ell b^{\frac 32\ell }}{\left( 1-b^\ell
\right) ^3}\sum_{j=1}^{\ell -1}\exp \left( -\frac{\hbar q^2}{2mw}\frac{%
\left( 1-b^j\right) \left( 1-b^{\ell -j}\right) }{1-b^\ell }\right) +\left[
\sum_{\ell =1}^\infty \frac{\xi ^{\ell -1}u^\ell b^{\frac 32\ell }}{\left(
1-b^\ell \right) ^3}\exp \left( -\frac{\hbar q^2}{4mw}\frac{1+b^\ell }{%
1-b^\ell }\right) \right] ^2\right\} .
\end{equation}

Using $\left( \sum_{\ell =1}^\infty a_\ell \right) ^2=\sum_{\ell =2}^\infty
\sum_{j=1}^{\ell -1}a_ja_{\ell -j}$ and defining
\begin{equation}
Q_{\ell ,j}\left( b\right) =\frac{1-b^\ell }{\left( 1-b^j\right) \left(
1-b^{\ell -j}\right) },
\end{equation}
the terms can be combined into
\begin{equation}
{\cal G}_2\left( u,{\bf q}\right) =\Xi _I\left( u\right) \sum_{\ell
=2}^\infty \frac{\xi ^{\ell -1}u^\ell b^{\frac 32\ell }}{\left( 1-b^\ell
\right) ^3}\sum_{j=1}^{\ell -1}\left( \exp \left( -\frac{\hbar q^2}{2mw}%
\frac 1{Q_{\ell ,j}\left( b\right) }\right) +\xi \left( Q_{\ell ,j}\left(
b\right) \right) ^3\exp \left( -\frac{\hbar q^2}{2mw}Q_{\ell ,j}\left(
b\right) \right) \right) .
\end{equation}
It should be noticed that only cycles with length at least two contribute to
the pair correlation function, as is to be expected. Because the series
expansion of ${\cal G}_2\left( u,{\bf q}\right) $ in powers of $u$ yields $%
{\Bbb Z}_I\left( N\right) Ng_{{\bf q}}$ as the coefficient of $u^N$ one
obtains immediately
\begin{equation}
Ng_{{\bf q}}=\frac 1{{\Bbb Z}_I\left( N\right) }\sum_{\ell =2}^N{\Bbb Z}%
_I\left( N-\ell \right) \frac{\xi ^{\ell -1}b^{\frac 32\ell }}{\left(
1-b^\ell \right) ^3}\sum_{j=1}^{\ell -1}\left( \exp \left( -\frac{\hbar q^2}{%
2mw}\frac 1{Q_{\ell ,j}\left( b\right) }\right) +\xi \left( Q_{\ell
,j}\left( b\right) \right) ^3\exp \left( -\frac{\hbar q^2}{2mw}Q_{\ell
,j}\left( b\right) \right) \right) .
\end{equation}
In the case of the pair correlation function $g_{{\bf q}}$ the sum rule $g_{%
{\bf q}=0}=N-1$ can also be checked. The proof proceeds by induction, but it
is rather tedious and is omitted here. All details on this calculation are
provided upon request.

\section{Boson density and pair correlation function}

In this section the density and the pair correlation of the model are
evaluated for the boson case, and the condensation effects on these
quantities are studied.

\subsection{Density}

The density $n\left( r\right) $ in the case of boson statistics can be
obtained from $n_{{\bf q}}$ and reads:
\begin{equation}
n\left( r\right) =\frac 1N\sum_{\ell =1}^N\frac{{\Bbb Z}_I\left( N-\ell
\right) }{{\Bbb Z}_I\left( N\right) }\frac{\xi ^{\ell -1}b^{\frac 32\ell }}{%
\left( 1-b^\ell \right) ^3}\left( \frac{wm}{\pi \hbar }A_\ell \right)
^{3/2}\exp \left( -\frac{mwr^2}\hbar A_\ell \right) ,  \label{eqn_r}
\end{equation}
with
\begin{equation}
A_\ell =\frac 1{\coth \frac 12\ell \beta \hbar w+\frac 1N\left( \frac w\Omega
\coth \frac 12\beta \hbar \Omega -\coth \frac 12\beta \hbar w\right) },
\end{equation}
where $r$ stands for the distance from the center of the confining
potential. The density is centrosymmetric, as a consequence of the isotropy
of the model. Introducing the activity $\rho _j$ as in I from ${\Bbb Z}%
_B\left( N\right) =\frac{b^{\frac 32N}}{\prod_{j=1}^N\left( 1-b^j\right) ^3}%
\prod_{j=0}^N\rho _j$ the density can be rewritten as follows
\begin{equation}
n\left( r\right) =\frac 1N\sum_{\ell =1}^N\frac 1{\left( 1-b^\ell \right) ^3}%
\left( \frac{wmA_\ell }{\pi \hbar }\right) ^{3/2}\exp \left( -\frac{mw}\hbar
r^2A_\ell \right) \prod_{j=N-\ell +1}^N\frac{\left( 1-b^j\right) ^3}{\rho _j}%
,
\end{equation}
which allows for a recursively defined expression well suited for numerical
evaluation
\begin{equation}
n\left( r\right) =\frac 1N\frac{\left( 1-b^N\right) ^3}{\rho _N}\left(
\!\!a_1+\frac{\left( 1-b^{N-1}\right) ^3}{\rho _{N-1}}\left( \!\!a_2+\frac{%
\left( 1-b^{N-2}\right) ^3}{\rho _{N-2}}\left( \!\!a_3+...+\frac{\left(
1-b^2\right) ^3}{\rho _2}\left( \!\!a_{N-1}+\frac{\left( 1-b\right) ^3}{\rho
_1}a_N\right) \!\!\right) \!\!\right) \!\!\right)
\end{equation}
with
\begin{equation}
a_\ell =\left( \frac{wm}{\pi \hbar }\right) ^{3/2}\frac{A_\ell ^{3/2}}{%
\left( 1-b^\ell \right) ^3}e^{-\rho ^2A_\ell }  \label{al}
\end{equation}
where $\rho =r\sqrt{\frac{mw}\hbar }$ is a natural dimensionless quantity
proportional to the distance from the center. Since $T=tT_c\sim tN^{1/3}$
(where $T_c$ is the condensation temperature for the Bose-Einstein
transition) $\rho /N^{1/6}$ is a natural quantity against which to plot the
density. The results are summarized in two figures. In Fig. 1, the density $%
n\left( 0\right) /n_{T=0}\left( 0\right) $ in the origin is shown (where $%
n_{T=0}\left( 0\right) $ is the density in the origin at zero temperature),
and exhibits a pronounced dependence on the condensation temperature. In
Fig. 2, $n\left( r\right) /n\left( 0\right) $ is plotted as a function of $r$
for 1000 particles. For comparison, the corresponding densities for the case
of distinguishable particles are plotted in Fig. 3. $T_c$ is only used as a
reference temperature for comparison purposes to Fig. 2; it does not have
the meaning of a condensation temperature if the particles are
distinguishable.

The typical shape of the density as a function of the temperature is shown
in Fig. 4 for $T=0,$ $T=0.9T_c,$ $T=T_c$ and $T=1.1T_c,$ where the spatial
dependence of the density $n\left( x,0,z\right) $ is plotted at a fixed
value $y=0$ for 1000 particles. It should be noted that the sudden
appearance of an intense peak below $T_c$ when sweeping through the
condensation temperature is manifestly present also in isotropic systems.

The center-of-mass contribution to the density can be substantial for a
limited number of particles. For 1000 particles this single degree of
freedom quantitatively has a negligible contribution to the density as a
function of $r\sqrt{w};$ the effects of the interaction enter in the
eigenfrequency $w=\sqrt{\Omega ^2-N\omega ^2}$ which determines the scaling
parameters in the figures.

\subsection{Pair correlation function}

An analogous analysis as for the density can be made for the pair
correlation function:
\begin{equation}
g\left( r\right) =\frac 1N\sum_{\ell =2}^N\frac{{\Bbb Z}_I\left( N-\ell
\right) }{{\Bbb Z}_I\left( N\right) }\frac{\xi ^{\ell -1}b^{\frac 32\ell }}{%
\left( 1-b^\ell \right) ^3}\sum_{j=1}^{\ell -1}\left( \frac{mw}{2\pi \hbar }%
Q_{\ell ,j}\right) ^{3/2}\left[ \exp \left( -\frac{mwr^2}{2\hbar }Q_{\ell
,j}\right) +\xi \exp \left( -\frac{mwr^2}{2\hbar }\frac 1{Q_{\ell ,j}}%
\right) \right] .  \label{eqg_r}
\end{equation}
Introducing
\begin{equation}
q_\ell =\sum_{j=1}^{\ell -1}\left( \frac{mw}{2\pi \hbar }Q_{\ell ,j}\right)
^{3/2}\frac 1{\left( 1-b^\ell \right) ^3}\left[ \exp \left( -\frac{mwr^2}{%
2\hbar }Q_{\ell ,j}\right) +\exp \left( -\frac{mwr^2}{2\hbar }\frac 1{%
Q_{\ell ,j}}\right) \right] ,
\end{equation}
(with $q_1=0)\ $a recurrence relation for $g\left( r\right) $ can be
obtained similarly as for $n\left( r\right) ,$ but with $a_\ell $ from (\ref
{al}) replaced by $q_\ell .$ In the same units as for the density, $g\left(
r\right) /g\left( 0\right) $ is plotted in Fig. 5. Similarly as for the
density in the previous subsection, $g\left( r\right) /g\left( 0\right) $ of
distinguishable particles is shown in Fig. 6 for comparison to the boson
case in Fig. 5.

The interpretation of Fig. 5 requires some caution, because $g\left(
r\right) /g\left( 0\right) $ is plotted, and the magnitude of $g\left(
0\right) $ strongly depends on the condensation temperature. Nevertheless,
it is clearly seen that the probability to find another particle at a
relatively small distance $r$ from some particle is very pronounced in the
condensate. Above the critical temperature a more substantial contribution
is obtained at relatively large distances, but the boson character still
manifests itself by a larger probability to find particles at relatively
short distances from the center.

\section{Discussion and conclusions}

The present paper concludes the boson part of a study of an exactly soluble
model containing one type of particles. The system exhibits condensation at
a finite temperature $T_c$. The thermodynamical quantities such as the
internal energy, the specific heat and the moment of inertia have been
studied before by the present authors and also by others \cite
{Grossman,Grossman2,Ketterle,Kirsten,Haugerud} as far as some
non-interacting aspects or ground state properties \cite{CohenLee} are
concerned. The density of this model is an important response property.
Precisely the concentration variations as a function of the cooling and the
field are invoked to establish the condensation transition \cite{BEC1,BEC2}.
Therefore it is comforting that the predictions of this theoretical model
--which of course constitutes a simplification-- bare some resemblance to
the simulated density of an anisotropic boson oscillator model \cite{Krauth}
and with the experimental situation in several aspects. It should be noted
(i) that the magnetically induced anisotropy of the trap is not taken into
account in the present paper, and (ii) that the interparticle interactions
are replaced by harmonic two-body interactions. Furthermore it is assumed
that the time scale of the experiment allows for an interpretation in terms
of the thermal equilibrium response properties. It is clear that these
simplifications deserve further investigations.

Also the pair correlation function of the model is an important quantity
especially if one wants to investigate the modifications due to a more
realistic interparticle interaction using a variational approach. Indeed, an
estimate of the effective interparticle potential along the lines of \cite
{FeynHibbs,FeynKlei} requires the pair correlation function of this exactly
soluble model, that is taken as the zeroth order approximation to the system.

From the methodological point of view, the projection and the generating
function technique allows to obtain tractable and exact expressions for the
thermodynamic quantities and for the static response functions of a system
of identical boson oscillators.

\acknowledgments
Part of this work is performed in the framework of the FWO projects No.
2.0093.91, 2.0110.91, G. 0287.95 and WO.073.94N (Wetenschappelijke
Onderzoeksgemeenschap, Scientific Research Community of the FWO on
``Low-Dimensional Systems''), and in the framework of the European Community
Program Human Capital and Mobility through contracts no. CHRX-CT93-0337 and
CHRX-CT93-0124. One of the authors (F.B.) acknowledges the FWO (Fund for
Scientific Research-Flanders) for financial support.

\begin{center}
{\bf Figure captions}
\end{center}

\begin{description}
\item[Fig. 1:]  Boson density $n\left( 0\right) /n_{T=0}\left( 0\right) $ at
the origin as a function of temperature.

\item[Fig. 2:]  Scaled density of bosons $n\left( r\right) /n\left( 0\right)
$ for 1000 particles as a function of the distance $r$ from the center for
several temperatures.

\item[Fig. 3:]  Scaled density $n\left( r\right) /n\left( 0\right) $ for
1000 distinguishable particles for comparison to Fig. 2.

\item[Fig. 4:]  Probability density $n\left( x,0,z\right) $ for 1000
particles as a function of $x$ and $z$ for $T=0,$ $T=0.9T_c,$ $T=T_c$ and $%
T=1.1T_c$.

\item[Fig. 5:]  Scaled pair correlation function $g\left( r\right) /g\left(
0\right) $ of 1000 bosons as a function of the distance $r$ for several
temperatures.

\item[Fig. 6:]  Scaled pair correlation function $g\left( r\right) /g\left(
0\right) $ of 1000 distinguishable particles for comparison to Fig. 5.
\end{description}

\end{document}